\documentclass[pra,nofootinbib]{revtex4-2}

\usepackage{amsmath, amssymb, amsthm, graphicx, epsfig, fancyhdr,epsfig, slashed}

\usepackage{tikzsymbols}
\usepackage{natbib}
\usepackage{float}

\newcommand{\sgn}{\operatorname{sgn}}
\newtheorem{lem}{Lemma}
\newtheorem{thm}{Theorem}

\begin{document}

\title{Quantized Noncommutative Riemann Manifolds and Stochastic
Processes: The theoretical foundations of the square root of Brownian motion}


%

%
%


\author{Marco Frasca}
\email{marcofrasca@mclink.it}
\affiliation{Via Erasmo Gattamelata, 3 \\ 00176 Roma (Italy)}

\author{Alfonso Farina}
\email{alfonso.farina@outlook.it}
\affiliation{Via Helsinki, 14 \\ 00144 Roma (Italy)}

\author{Moawia Alghalith}
\email{malghalith@gmail.com}
\affiliation{Department of Economics, The University of the West Indies, St.
Augustine (Trinidad and Tobago)} 

%
%
%
%

\begin{abstract}
We lay the theoretical and mathematical foundations of the
square root of Browniam motion and we prove the existence of such a process.
In doing so, we consider Brownian motion on quantized noncommutative
Riemannian manifolds and show how a set of stochastic processes on sets of
complex numbers can be devised. This class of stochas-tic processes are
shown to yield at the outset a Chapman-Kolmogorov equation with a
complex diffusion coefficient that can be straightforwardly reduced to the
Schr\"odinger equation. The existence of these processes has been
recently shown numerically. In this work we provide an analogous support for
the existence of the Chapman-Kolmogorov-Schr\"odinger equation for
them, performing a Monte Carlo study. It is numerically seen as a Wick
rotation can turn the heat kernel into the Schr\"odinger one, mapping
such kernels through the corresponding stochastic processes. In this way, we
introduce a new kind of improper complex stochastic process. This permits a
reformulation of quantum mechanics using purely geometrical concepts that
are strongly linked to stochastic processes. Applications to economics are
also entailed.
\end{abstract}

\maketitle

\section{Introduction}

\label{sec1}

One of the hotly debated problems about quantum mechanics is if it could be
derived from some stochastic process. One the most promising proposal was
put forward by Nelson \cite{nels}. This idea was widely discussed \cite%
{Guerra:1981ie,hang,skor,bla1,bla2,bla3} but it remains an open question if
it could be a solution to the problem. Quite recently, we proposed an
approach, based on the extraction of the square root of a Wiener process %
\cite{fari,fra1}, that identifies a new class of stochastic processes based
on complex evaluated random variables\footnote{%
These come out naturally trying to extend the Tartaglia-Pascal triangle to
quantum mechanics. As shown in \cite{fari}, the correspondence is between
the binomial coefficient $\binom{n}{k}$ and the discrete quantum analog $%
\sqrt{\binom{n}{k}}2^{-\frac{n}{2}}e^{i\left[\frac{(k-n/2)^2}{n}\sqrt{\frac{n%
}{4}-1}-\frac{1}{2}\arctan\sqrt{\frac{n}{4}-1}\right]}$.}. These processes
can have a Schr\"odinger equation as Kolmogorov-Chapman diffusion equation.
This proposal was put at test with a numerical study in \cite{fra2} and we
showed a theorem about fractional powers of Wiener processes. Using a simple
integration technique, the Euler-Maruyama method, we solved the stochastic
differential equations arising in the square root case proving the existence
of such a process. Complex evaluated stochastic processes can have
convergence problems when managed with the standard techniques of real
evaluated stochastic processes. This is the main reason why we recurred to
numerical methods. Particularly, in this paper we will give numerical
evidence of the existence of the corresponding diffusion process, given by
the Schr\"odinger equation, for the square root case. 

These processes arise naturally as a Brownian motion on a noncommutative
Riemann manifold. Connes, Chamseddine and Mukhanov proved that such a
noncommutative manifold is quantized and made by two kinds of elementary
volumes \cite{Chamseddine:2014nxa,Chamseddine:2014uma}, identified by the
unities $(1,i)$, and it is from here that the deep connection with
stochastic processes starts. This means that the relation between an
ordinary diffusion process \textsl{a la} Fourier and the Schr\"odinger
equations, formally given by a Wick rotation $t\rightarrow -it$, has a deep
physical meaning. Mathematically, as already said, it entails the
introduction of a new class of stochastic processes: the fractional powers
of a Wiener process \cite{fari,fra1,fra2}. This connection with the
noncommutative geometry is a natural one as a square root stochastic process
can only be built if a Clifford algebra \cite{naka} exists to support it.
Otherwise, the ordinary Wiener process cannot be recovered by taking the
square because a spurious shifting term will appear.

Although there is a wide literature on stochastic processes in
noncommutative geometry (e.g. see \cite{con1,con2}), the aim of this paper
is to present and prove the existence of a new class of stochastic processes
that could have a possible application in noncommutative geometry as
discussed in the recent works by Connes, Chamseddine and Mukhanov \cite%
{Chamseddine:2014nxa,Chamseddine:2014uma}. Our conclusions do not need
noncommutative geometry to hold. 

This class of stochastic processes can be classified as improper complex
stochastic processes \cite{sch}. We use them to obtain a reformulation of
quantum mechanics starting from noncommutative geometry and its deep
connection with stochastic processes. This implicate an understanding of
quantum mechanics as a motion on a quantized manifold. On a similar ground,
relevant applications to economy and finance are also expected \cite{alg}.

In this paper, \textbf{we will introduce the theoretical and mathematical} 
\textbf{foundations of the square root of Browniam motion and we prove the
existence of such a process. We will also} show, by numerical evidence, that
the square root process gives rise to a diffusion process ruled by a Schr%
\"{o}dinger-like equation. We also show that this is a complex valued
stochastic process. The work is structured as follows. In Sec.~\ref{sec2},
we give some elements of noncommutative geometry for a quantized Riemann
manifold and introduce the stochastic process on it. In Sec.~\ref{sec3}, we
yield some results about fractional powers of a Wiener process, specifically
for the power $1/2$. In Sec.~\ref{sec3a}, we present a theorem showing that
the square root process in indeed a complex valued stochastic process. In
Sec.~\ref{sec4}, we present the results of our Monte Carlo study of the
diffusion process for this class of stochastic processes. Finally, in Sec.~%
\ref{sec5}, conclusions are given.

\section{Quantized Riemann manifolds}

\label{sec2}

\subsection{Noncommutative geometry}

A noncommutative geometry is characterized by the triple $(\mathcal{A},H,D)$
being $\mathcal{A}$ a set of operators forming a $^*$-algebra, $H$ a Hilbert
space and $D$ a Dirac operator. This yields that the volume of the
corresponding noncommutative Riemann manifold is quantized with two distinct
classes of unity of volume $(1,i)$. A proof of this theorem was provided by
Connes, Chamseddine and Mukhanov\cite%
{Chamseddine:2014nxa,Chamseddine:2014uma}. The need of two kinds of
elementary volumes arises from the fact that the Dirac operator should not
be limited to Majorana (neutral) states in the Hilbert space but we have
more general states and we have to add a charge conjugation operator $J$ to
our triple. Finally, we recall that the Clifford algebra of Dirac matrices
implies the existence of a $\gamma^5$ matrix \cite{cliff}, the chirality
matrix that changes the parity of the states. For a commutative Riemann
manifold, the algebra $\mathcal{A}$ is the Abelian algebra of smooth
functions. One has $[D,a]=i\gamma\cdot\partial a$, and noting that, in four
dimensions, $x_1,\ x_2,\ x_3,\ x_4$ are legal functions of $\mathcal{A}$, we
can generate $\gamma^5$ as $[D,x_1][D,x_2][D,x_3][D,x_4]=\gamma^1\gamma^2%
\gamma^3\gamma^4=-i\gamma^5$. Similarly, for arbitrary functions in $%
\mathcal{A}$, $a_0,\ a_1,\ a_2,\ a_3,\ a_4,\ \ldots\ a_d$, summing over all
the possible permutations one has a Jacobian. Then, we can define a more
general chirality operator 
\begin{equation}
\gamma=\sum_P(a_0[D,a_1]\ldots[D,a_d]),
\end{equation}
that, in four dimension, gives 
\begin{equation}
\gamma=-iJ\cdot\gamma^5=-i\cdot\mathrm{det}(e)\gamma^5
\end{equation}
being $J$ the Jacobian, $e^a_\mu$ the vierbein \cite{naka} for the Riemann
manifold, characterizing the metric, and $\gamma^5=i\gamma^1\gamma^2\gamma^3%
\gamma^4$ for $d=4$, a well-known result. We used the fact that $\mathrm{det}%
(e)=\sqrt{g}$, being $g_{\mu\nu}$ the metric tensor. So, our definition of
chirality operator is just proportional to the metric factor that yields the
volume of a Riemannian orientable manifold.

A Riemannian manifold can be properly quantized when, instead of functions,
we consider operators $Y$ belonging to an operator algebra $\mathcal{%
A^{\prime}}$. These operators have the properties 
\begin{equation}  \label{eq:Y}
Y^2=\kappa I \qquad Y^\dagger=\kappa Y.
\end{equation}
These are operators that have the role of coordinates as in the Heisenberg
commutation relations. To account for the existence of the conjugation of
charge operator $C$ such that $CAC^{-1}=Y^\dagger$, we need two sets of
coordinates, $Y_+$ and $Y_-$ as we expect a conjugation of charge operator $C
$ to exist such that $CAC^{-1}=Y^\dagger$. This is the analogous of complex
conjugation for a function. Such coordinates appear naturally out of a Dirac
algebra of gamma matrices. Indeed, a natural way to write down the operators 
$Y$ is by using a Clifford algebra of Dirac matrices $\Gamma^A$ such that 
\begin{equation}
\{\Gamma^A,\Gamma^B\}=2\delta^{AB}, \qquad (\Gamma^A)^*=\kappa\Gamma^A
\end{equation}
with $A,B=1\ldots d+1$, so that 
\begin{equation}
Y=\Gamma^AY^A.
\end{equation}
We will need two different sets of gamma matrices for $Y_+$ and $Y_-$ having
these independent traces. Using the charge conjugation operator $C$, we can
introduce a new coordinate 
\begin{equation}
Z=2ECEC^{-1}-I
\end{equation}
where $E=(1+Y_+)/2+(1+iY_-)/2$ is a projector for the coordinates. It is not
difficult to see that the spectrum of $Z$ is the set $(1,i)$, given eq.(\ref%
{eq:Y}). We can now generalize our definition of the chirality operator by
taking the trace on $Z$s, properly normalized to the number of components.
This yields 
\begin{equation}  \label{eq:Z}
\frac{1}{n!}\langle Z[D,Z]\ldots[D,Z]\rangle=\gamma,
\end{equation}
being the average $\langle\ldots\rangle$, in this case, just matrix traces.
We can now see the quantization of the volume. Let us consider a three
dimensional manifold $M$ and the sphere $\mathbb{S}^2$. From eq.(\ref{eq:Z})
one has 
\begin{equation}
V_M=\int_M\frac{1}{n!}\langle Z[D,Z]\ldots[D,Z]\rangle d^3x.
\end{equation}
By taking the traces we get 
\begin{equation}
V_M=\int_M\left(\frac{1}{2}\epsilon^{\mu\nu}\epsilon_{ABC}Y^A_+\partial_\mu
Y^B_+\partial_\nu Y^C_++ \frac{1}{2}\epsilon^{\mu\nu}\epsilon_{ABC}Y^A_-%
\partial_\mu Y^B_-\partial_\nu Y^C_-\right)d^3x.
\end{equation}
It is not difficult to see that this will reduce to~\cite%
{Chamseddine:2014nxa,Chamseddine:2014uma} 
\begin{equation}
\mathrm{det}(e^a_\mu)=\frac{1}{2}\epsilon^{\mu\nu}\epsilon_{ABC}Y^A_+%
\partial_\mu Y^B_+\partial_\nu Y^C_++ \frac{1}{2}\epsilon^{\mu\nu}%
\epsilon_{ABC}Y^A_-\partial_\mu Y^B_-\partial_\nu Y^C_-.
\end{equation}
The coordinates $Y_+$ and $Y_-$ belong to unitary spheres while the Dirac
operator has a discrete spectrum as it is defined on a compact manifold.
This means that we are covering all the manifold with a large integer number
of these spheres. Therefore, the volume is quantized as this is required by
the above condition. An extension to four dimensions is also possible with
some more work \cite{Chamseddine:2014nxa,Chamseddine:2014uma}.

\subsection{Stochastic processes on a quantized manifold}

\begin{figure}[H]
\centerline{\includegraphics[height=150pt]{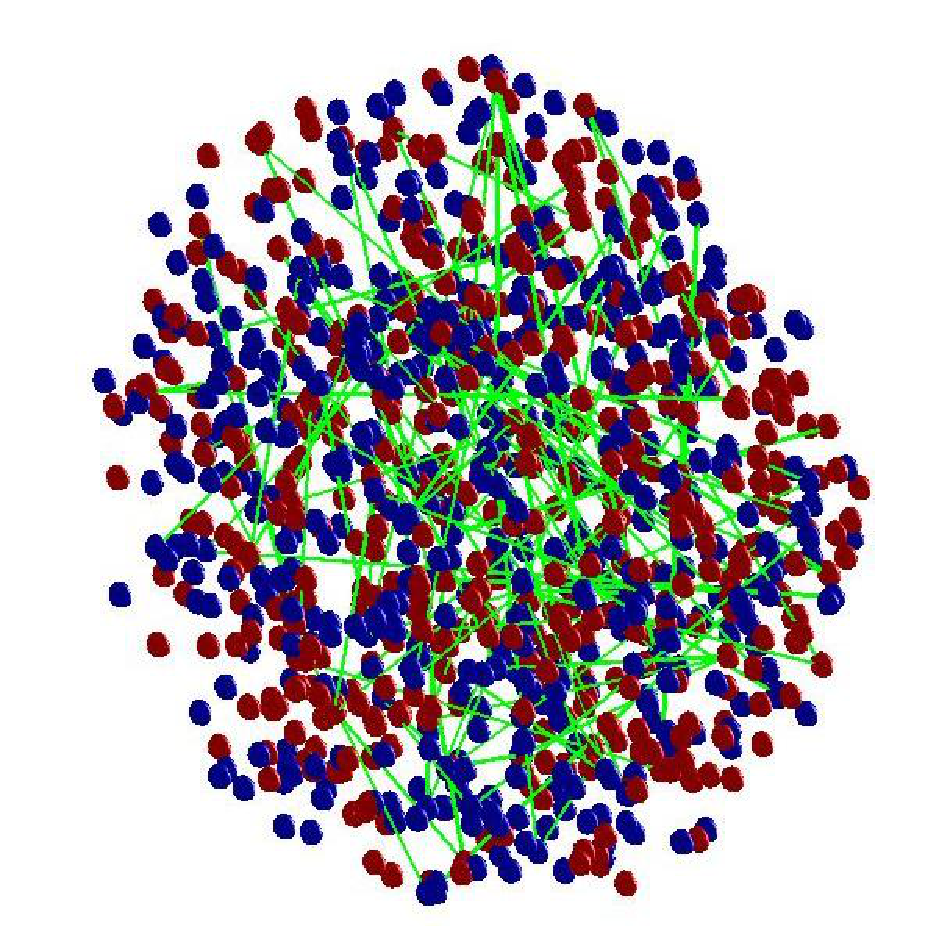}}
\caption{Pictorial representation of motion (green lines) of a particle on a
noncommutative Riemann manifold. }
\label{fig:fig1}
\end{figure}

We expect that a Wiener process on a quantized manifold will account for the
spectrum $(1,i)$ of the coordinates on the two kinds of spheres $Y_+,\ Y_-$.
Assuming a completely random distribution of the two kinds of spheres that
make the Riemann manifold, the result will depend on the motion of the
particle on it. A process $\Phi$ can be defined such that, like for tossing
of a coin, one gets either $1$ or $i$ as outcome, once we assume that the
distribution of the unitary volumes is uniform. The definition of this
process is 
\begin{equation}  \label{eq:Phi}
\Phi = \frac{1+B}{2}+i\frac{1-B}{2}
\end{equation}
with $B$ a Bernoulli process such that $B^2=I$ that yields the value $\pm 1$
depending on the unitary volume hit by the particle. It is also $\Phi^2=B$.
For a Brownian motion of the particle on such a manifold, the possible
outcomes will be either $Y_+$ or $Y_-$. For a given set of $\Gamma$ matrices
and chirality operator $\gamma$, one can write the most general form for
such a stochastic process as (summation on $A$ is implied) 
\begin{equation}  \label{eq:ncgSP}
dY=\Gamma^A\cdot (\kappa_A+\xi_A dX_A\cdot B_A+\zeta_A
dt+i\eta_A\gamma^5)\cdot\Phi_A
\end{equation}
being $\kappa_A,\ \xi_A,\ \zeta_A,\ \eta_A$ arbitrary coefficients of this
linear combination (a pictorial view is given in Fig.~\ref{fig:fig1}). 
The Bernoulli processes $B_A$ and the Wiener process $dX_A$ are not
independent. We expect that the sign arising from the Bernoulli process
should be the same of that of the corresponding Wiener process. This
stochastic differential equation is the equivalent of the eq.(\ref{eq:Y})
for the coordinates on the manifold. As we will see below, this is the same
as the formula for the square root of a Wiener process. This represents the
motion of a particle on a quantized noncommutative Riemann manifold. In this
way, the Schr\"odinger equation can be removed from the state of a postulate
and, underlying quantum mechanics, we have a quantized manifold.

\section{Fractional powers of Wiener processes}

\label{sec3}

The first step is to prove the existence of the square root process. This
was already accomplished in \cite{fra2} using numerical techniques. Anyway,
we give the following theorem here:

\begin{thm}
Given a random variable $X(t)\sim N(0,\sqrt{t})$, on a time sequence $%
\{t(m)\}\in\mathbb{R}$ and $m\in\mathbb{N}$, the sequence $Y(m+1)=Y(m)+\sqrt{%
X(m)}$ exists and belongs to $\mathbb{C}$. Then, $Y(t)$ on the same
sequence, is a stochastic process representing the square root of a Brownian
motion.
\end{thm}

\begin{proof}
Let us write $X(m)=|X(m)|e^{ik(m)\pi}$, being $\{k(m)\}$ a random sequence in $\mathbb{N}$ corresponding to $\sgn(X(m))$. Then, $\sqrt{X(m)}=\sqrt{|X(m)|}e^{ik(m)\frac{\pi}{2}}\in\mathbb{C}$ , exists and is well defined. Then, also the sequence $Y(m+1)=Y(m)+\sqrt{X(m)}$ exists and is well-defined and is in $\mathbb{C}$.

But the sequence $\{Y(i)\}$ represents the stepwise solution, through the Euler-Maruyama method, of the stochastic equation
\begin{equation}
    dY=\sqrt{dX}
\end{equation}
being $dX$ a Brownian process by construction. Now, being the Brownian process continuous, the limit of the time step $\Delta t=t(m+1)-t(m)\rightarrow 0$ also exists and so, the square root process exists as well.
\end{proof}

With It\=o calculus we can express the ``square root'' process through more
elementary stochastic processes \cite{okse}, $(dW)^2=dt$, $dW\cdot dt=0$, $%
(dt)^2=0$ and $(dW)^\alpha=0$ for $\alpha>2$, we set 
\begin{equation}  \label{eq:sqrt}
dX=(dW)^\frac{1}{2}\overset{?}{=}\left(\mu_0+\frac{1}{2\mu_0}dT-\frac{1}{%
8\mu_0^3}dt\right)\cdot\Phi_\frac{1}{2}
\end{equation}
being $dT=dW\cdot\operatorname{sgn}(dW)$ a Tanaka process \cite{oks} such that $%
(dT)^2\sim(dW)^2\sim dt$, $\mu_0\ne 0$ an arbitrary scale factor and 
\begin{equation}
\Phi_\frac{1}{2}=\frac{1-i}{2}\operatorname{sgn}(dW)+\frac{1+i}{2}
\end{equation}
a Bernoulli process equivalent to a coin tossing that has the property $%
(\Phi_\frac{1}{2})^2=\operatorname{sgn}(dW)$. The possible outcomes for this process
are $1$ and $i$ and represent a particle executing Brownian motion
scattering two different kinds of small pieces of space, each one
contributing either 1 or i to the process, randomly. We have already seen
this process for the noncommutative geometry in eq.(\ref{eq:Phi}). We have
introduced the process $\operatorname{sgn}(dW)$ that yields just the signs of the
corresponding Wiener process. But Eq.(\ref{eq:sqrt}) is not satisfactory
for, taking the square, yields 
\begin{equation}
(dX)^2=\mu_0^2\operatorname{sgn}(dW)+dW
\end{equation}
and we do not exactly recover the original Wiener process. We see that we
have added a process that has an overall effect to shift upward the original
Brownian motion even if its shape is preserved.

This problem can be fixed by using the Clifford algebra formed by the Pauli
matrices \cite{cliff}. Taking two different Pauli matrices $\sigma_i,\
\sigma_k$ with $i\ne k$ such that $\{\sigma_i,\sigma_k\}=0$ we can rewrite
the above identity as 
\begin{equation}
I\cdot dX=I\cdot(dW)^\frac{1}{2}=\sigma_i\left(\mu_0+\frac{1}{2\mu_0}dT-%
\frac{1}{8\mu_0^3}dt\right)\cdot\Phi_\frac{1}{2}+i\sigma_k\mu_0\cdot\Phi_%
\frac{1}{2}
\end{equation}
and so, $(dX)^2=dW$ as it should. This idea can be easily generalized to
higher dimensions using Dirac's $\gamma$ matrices. 
We see that we have recovered a similar stochastic process as in eq.(\ref%
{eq:ncgSP}).

This view agrees very well with the recent results by Connes, Chamseddine
and Mukhanov \cite{Chamseddine:2014nxa,Chamseddine:2014uma} and yields a
hint for the underlying possible quantization of space.

We notice from this result that already the presence of the Tanaka process,
that is defined in a weak sense \cite{oks}, means that the square root
process is a complex valued stochastic process not in a proper sense \cite%
{sch}. We will see this below.

For consistency reasons, we also provide the operational definitions for the
involved processes needed to complete the above derivation. These are \cite%
{fra2} 
\begin{equation}
\operatorname{sgn}(dW)=\{\operatorname{sgn}(W_0),\operatorname{sgn}(W_1),\operatorname{sgn}(W_2),\ldots\}
\end{equation}
such that $(\operatorname{sgn}(dW))^2=I$, 
\begin{equation}
|dW| = \{|W_0|,|W_1|,|W_2|,\ldots\}
\end{equation}
and for the Tanaka process 
\begin{equation}
dT=|dW|\operatorname{sgn}(dW)=\{|W_0|\cdot\operatorname{sgn}(W_0),|W_1|\cdot\operatorname{sgn}%
(W_1),|W_2|\cdot\operatorname{sgn}(W_2),\ldots\}=dW.
\end{equation}
These definitions are also used in the numerical evaluation for the proof by
construction in Sec.~\ref{sec4}.

We can consider a more general ``square root'' process by adding a term
proportional to $dt$. We take for granted that the Pauli matrices are used
to remove the $\operatorname{sgn}$ so, we will permit us to neglect it. Assuming for
the sake of simplicity $\mu_0=1/2$, one has 
\begin{equation}  \label{eq:sqrtW}
dX(t)=[dW(t)+\beta dt]^\frac{1}{2}= \left[\frac{1}{2}+dW(t)\cdot\operatorname{sgn}%
(dW(t))+(-1+\beta\operatorname{sgn}(dW(t)))dt\right]\Phi_{\frac{1}{2}}(t),
\end{equation}
being $\beta$ an arbitrary constant. From the Bernoulli process $\Phi_{\frac{%
1}{2}}(t)$ one gets 
\begin{equation}  \label{eq:musig}
\mu =-\frac{1+i}{2}+\beta\frac{1-i}{2}\qquad\sigma^2=2D=-\frac{i}{2}.
\end{equation}
The presence of a complex valued pseudo-variance $\sigma^2$ show that the
square root is an improper complex-valued process \cite{sch}. So, we have
the following lemma:

\begin{lem}
A square root stochastic process is an improper complex-valued stochastic
process.
\end{lem}

Therefore, we have a double Fokker--Planck equation for a free particle,
being the distribution function $\hat{\psi}$ complex valued, 
\begin{equation}
\frac{\partial \hat{\psi}}{\partial t}=\left( \frac{1+i}{2}-\beta \frac{1-i}{%
2}\right) \frac{\partial \hat{\psi}}{\partial X}-\frac{i}{4}\frac{\partial
^{2}\hat{\psi}}{\partial X^{2}}.  \label{eq:fpsch}
\end{equation}%
This result is not unexpected as, having complex random variables, we should
have a Fokker--Planck equation for the real part and another for the
imaginary part. The surprising result is that we get an equation strongly
resembling the Schr\"{o}dinger equation. We will see below that we are
really recovering quantum mechanics, by recovering the heat kernel from the
Monte Carlo simulation of the ``square root'' process, after a Wick rotation
the square root process.

\section{\protect\bigskip Proof of the existence of the square root of the
Brownian motion}

\label{sec3a}


In this section we will go deeper into the properties of the square root
process, starting from the following theorem:


\begin{thm}
The square root of a standard Brownian motion can be given by 
\begin{equation}
\sqrt{W\left( t\right) }=\frac{t^{\frac{1}{4}}}{2\left( E\Phi ^{4}\right) ^{%
\frac{1}{4}}}\left[ \Phi -\mid \Phi \mid +i\left( \Phi +\mid \Phi \mid
\right) \right] ,  \label{1}
\end{equation}%
where $\Phi $ is a real-valued continuous random variable, $E\Phi =0,$ $%
-\Phi \mid \Phi \mid \sim N\left( 0,E\Phi ^{4}\right) ,$ $i=\sqrt{-1,}$ and $%
t$ is time.
\end{thm}

\begin{proof}
First, \textit{in general}, the process $\frac{t^{\frac{1}{4}%
}}{2\left( E\Phi ^{4}\right) ^{\frac{1}{4}}}\left[ \Phi -\mid \Phi \mid
+i\left( \Phi +\mid \Phi \mid \right) \right] $ exists since it is a
(typical) complex random variable. To show that it is the square root of a
Brownian motion in particular, we use $\left( \ref{1}\right) $ to get
\[
W\left( t\right) =\left( \frac{t^{\frac{1}{4}}}{2\left( E\Phi ^{4}\right) ^{%
\frac{1}{4}}}\left[ \Phi -\mid \Phi \mid +i\left( \Phi +\mid \Phi \mid
\right) \right] \right) ^{2}=-\frac{\sqrt{t}}{\sqrt{E\Phi ^{4}}}\Phi \mid
\Phi \mid . 
\]%

Clearly, from the definition of a Brownian motion, the process $-\frac{\sqrt{%
t}}{\sqrt{E\Phi ^{4}}}\Phi \mid \Phi \mid $ is a standard Brownian motions ($%
E\left[ -\frac{\sqrt{t}}{\sqrt{E\Phi ^{4}}}\Phi \mid \Phi \mid \right] =0$
and $Var\left( -\frac{\sqrt{t}}{\sqrt{E\Phi ^{4}}}\Phi \mid \Phi \mid
\right) $ $=t)$, since a Brownian motion is defined as $\sqrt{t}X$, where $X$
is a Gaussian variable. 
\end{proof}

\bigskip

Similarly, using the same procedure, it can be shown that 
\[
\sqrt{dW\left( t\right) }=\frac{\left( dt\right) ^{\frac{1}{4}}}{2\left(
E\Psi ^{4}\right) ^{\frac{1}{4}}}\left[ \Psi -\mid \Psi \mid +i\left( \Psi
+\mid \Psi \mid \right) \right] , 
\]%
where $\Psi $ is a real-valued random variable so that $-\Psi \mid \Psi \mid
\sim N\left( 0,E\Psi ^{4}\right) .$

\bigskip

\textbf{Properties.}

\medskip

The square root of the Brownian motion has some of the typical properties of
a Brownian motion, such as


\begin{enumerate}
\item It is continuous almost surely.

This follows directly from the continuity of $\Phi.$

\item It is nowhere differentiable almost surely.


\begin{proof}
$d\sqrt{W\left( t\right) }=\frac{1}{2\sqrt{W\left(
t\right) }}dW\left( t\right) +....$ Thus,

$\frac{d\sqrt{W\left( t\right) }}{dt}=\frac{1}{2\sqrt{W\left( t\right) }}%
\frac{dW\left( t\right) }{dt}+...$
 and $W\left( t\right) $ is nowhere differentiable.
\end{proof}

This result is empirically verified in \cite{fra2}.

\item It starts from zero almost surely.

This directly follows from $W\left( 0\right) =0.$

\item Scaling: $c^{\frac{1}{4}}\sqrt{W\left( t/c\right) }$ $,c^{-\frac{1}{4}}%
\sqrt{W\left( ct\right) }$ and $\sqrt{W\left( t\right) }$ are square root
Brownian motions$,$ $c\neq 0.$ This follows directly from $\left( \ref{1}%
\right) .$

\item However, unlike a Brownian motion, 
\[
E\sqrt{W\left( t\right) }=\left( i-1\right) \frac{t^{\frac{1}{4}}}{2\left(
E\Phi ^{4}\right) ^{\frac{1}{4}}}E\mid \Phi \mid 
\]%
and 
\[
Var\left( \sqrt{W\left( t\right) }\right) =\frac{\sqrt{t}}{4\sqrt{E\Phi ^{4}}%
}\left[ Var\left( \Phi \right) +Var\left( \mid \Phi \mid \right) \right] .
\]
\end{enumerate}


\section{Monte Carlo study}

\label{sec4}

A recent Monte Carlo study by the authors \cite{fra2} has shown the
existence of fractional Wiener processes, provided a proper definition of
the involved random evaluated functions is given. In this way, a
straightforward numerical implementation is possible. Having this in mind,
we use the same technique to perform a Monte Carlo evaluation of the
diffusion process involved with our complex random processes and show that
the so obtained mean, variance and probability distribution agree fairly
well with what we have obtained theoretically so far. We just note that mean
and variance should be evaluated by dividing by $\mu_0$ and $\mu_0^2$
respectively. $\mu_0$ should be chosen greater than one. In our case $\beta$%
, that appears in eq.~(\ref{eq:sqrtW}), is assumed to be zero.

\begin{figure}[H]
\centerline{\includegraphics[height=150pt]{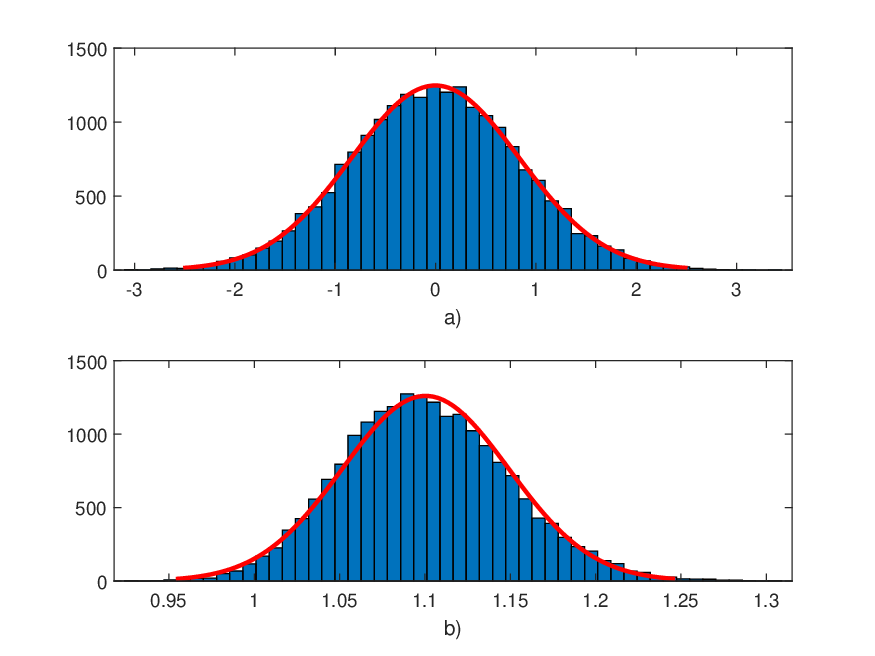}} 
\caption{Comparison between the heat kernel a) and the Schr\"odinger kernel
b) obtained after a Wick rotation.}
\label{fig:fig2}
\end{figure}

We performed a Monte Carlo study where each Brownian path is evaluated for
1000 steps for 20000 runs\footnote{%
The code is available on request to M.F.}. In this way we were able to
evaluate both the Wiener process, its square root and eq.~(\ref{eq:sqrtW})
obtained by Euler-Maruyama method. We expect that the kernel is the standard
heat kernel for the first case and a Schr\"odinger kernel otherwise. But
this should be correlated by a Wick rotation. So, in order to perform a fit
with a Gaussian distribution, we need to be certain that the phases of the
Schr\"odinger kernel, producing the imaginary part, are removed after a Wick
rotation. This is indeed the case. Therefore, given the set of random
complex numbers $\psi$ obtained by numerically evaluating the square of a
Wiener path sample, we evaluate the module $\rho$ and the phase $\theta$ for
each one of them. Now we have for the Schr\"odinger kernel 
\begin{equation}
\hat\psi=(4\pi it)^{-\frac{1}{2}}\exp{\left(ix^2/4t\right)}=(4\pi t)^{-\frac{%
1}{2}}\left[\cos\left(\frac{x^2}{4t}-\frac{\pi}{4}\right)+i\sin\left(\frac{%
x^2}{4t}-\frac{\pi}{4}\right)\right].
\end{equation}
A Wick rotation, $t\rightarrow -it$, turns it into a heat kernel giving
immediately 
\begin{equation}  \label{eq:wick}
K=(4\pi t)^{-\frac{1}{2}}\left[\cos\left(i\frac{x^2}{4t}-i\frac{\pi}{4}%
\right)+i\sin\left(i\frac{x^2}{4t}-i\frac{\pi}{4}\right)\right]e^{\frac{\pi}{%
4}}.
\end{equation}
Given the phases and modules computed by our set of samples, this can be
easily expressed using them. The result is given in Fig.~\ref{fig:fig2}. 
One sees that one gets a perfect normal distribution in both the cases as it
should. We just note that, in our case, the Schr\"odinger kernel has its
center shifted, in agreement with our expectations.

\begin{figure}[H]
\centerline{\includegraphics[height=150pt]{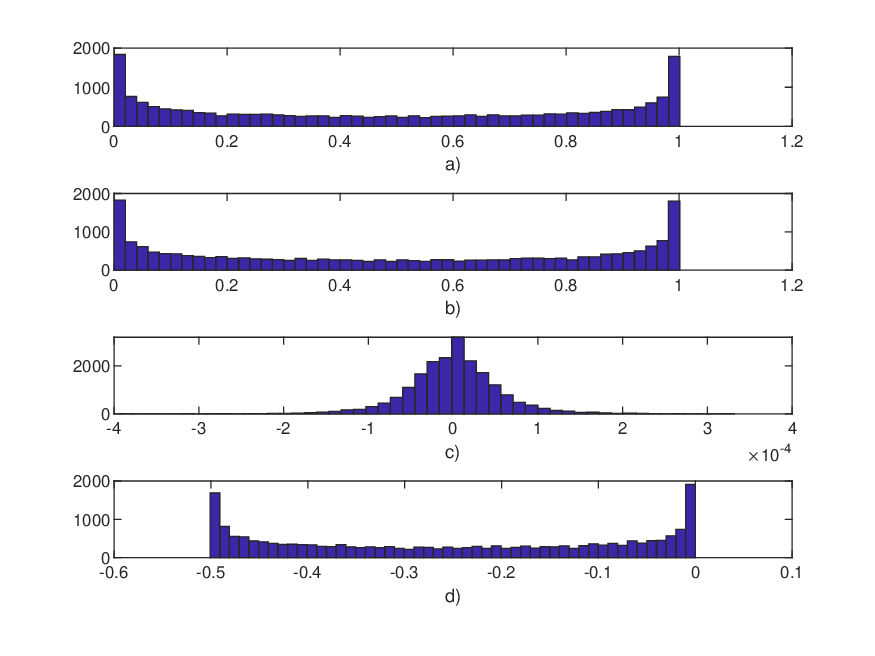}}
\caption{Distributions of the means (real part a) and imaginary part b)) and
variances (real part c) and imaginary part d)) of the square root process.}
\label{fig:fig3}
\end{figure}

In Fig.~\ref{fig:fig3}, we show the distributions of the averages and the
variances of the square root of the Wiener process. 
In table~\ref{tab:tab1}, we report the values of their means, variances and
diffusion coefficients. 
\begin{table}[tbp]
\caption{Means, variances and diffusion coefficients.}
\label{tab:tab1}%
\begin{tabular}{|l|c|c|c|}
\hline
Process & Mean & Variance & Diffusion coefficient \\ \hline
Brownian & $-0.001\pm 0.004$ & $0.1667\pm 0.0011$ & $0.2041\pm 0.0013$ \\ 
Square root & $0.4986\pm 0.0025+(0.5016\pm 0.0025)i$ & $0\pm 4\cdot
10^{-7}-(0.2491\pm 0.0013)i$ & $-(0.249\pm 0.001)i$ \\ \hline
\end{tabular}%
\end{table}
The agreement with our theoretical results, looking at eq.~(\ref{eq:fpsch}),
is exceedingly good confirming that we are observing a diffusion process
ruled by the Schr\"odinger equation arising from the square root of a Wiener
process. Particularly, we notice the values $(1+i)/2$ for the mean of the
square root process, in agreement with eq.~(\ref{eq:musig}) (in our
numerical study is $\beta=0$), and the variance being $-i/4$ in agreement
with the expected diffusion coefficient. This appears to be just
Wick-rotated with respect to the case of the heat equation.

\section{Conclusions}

\label{sec5}

\textbf{We have layed the theoretical and mathematical} \textbf{foundations
of the square root of Browniam motion and we proved the existence of this
process. }We also have shown, also through a Monte Carlo study, the
existence of a diffusion process, described by a Schr\"{o}dinger equation,
arising by taking the square root of an ordinary Brownian motion. We have a
complete agreement with the theoretical expectations. As a concluding
remark, we are pleased to note that our theory has recently been applied in
the field of stock exchange prediction as a refinement of the Black and
Scholes equation \cite{alg}. Therefore, this process will have many
applications in economics. For example, it can be used to model stochastic
volatility, stochastic interest rate and asset pricing, among others.
Stochastic volatility is becoming increasingly popular in economics and
econometrics. This is very timely, since econophysics is an emerging field.
This paper can strengthen and help shape this new field of study.

\textbf{A natural future extension of this paper is to introduce stochastic
integrals for the square root of a  Brownian motion and compare them to
Ito's integrals.}

\textbf{%
%
%
%
}

\end{document}